\documentclass[12pt]{article}
\usepackage{graphicx}

\newcommand\pubnumber{SNSN-323-63}
\newcommand\pubdate{\today}

\def\institute{School of Particles and Accelerators, Institute for Research in Fundamental
Sciences(IPM),\\
P. O. Box 19 56 83 66 81, Tehran, Iran}

\def\Title#1{\begin{center} {\Large #1 } \end{center}}
\def\Author#1{\begin{center}{ \sc #1} \end{center}}
\def\Address#1{\begin{center}{ \it #1} \end{center}}

\newcommand\pubblock{\rightline{\begin{tabular}{l} \pubnumber\\
         \pubdate  \end{tabular}}}
\newenvironment{Abstract}{\begin{quotation}  }{\end{quotation}}
\newenvironment{Presented}{\begin{quotation} \begin{center} 
             PRESENTED AT\end{center}\bigskip 
      \begin{center}\begin{large}}{\end{large}\end{center} \end{quotation}}

\def\beq{\begin{equation}}
\def\eeq#1{\label{#1}\end{equation}}
\def\eeqn{\end{equation}}

\def\beqa{\begin{eqnarray}}
\def\eeqa#1{\label{#1}\end{eqnarray}}
\def\eeqan{\end{eqnarray}}

\let\bar=\overbar

\def\Dslash{\not{\hbox{\kern-4pt $D$}}}
\def\dslash{\not{\hbox{\kern-2pt $\del$}}}

\def\mt{m_t}

\def\msb{{\bar{\ssstyle M \kern -1pt S}}}

\usepackage{commath}

\usepackage{xspace}
\usepackage[]{subfig}
\input{ptdr-definitions.sty}

 \newcommand{\invfb}{\fbinv}
\newcommand{\qcd}{\ensuremath{\mathrm{QCD}}\xspace}
  \newcommand{\QCD}{\qcd}
   \newcommand{\wjets}{\ensuremath{\mathrm{W}+jets}\xspace}

  \newcommand{\ttbar}{\ensuremath{\mathrm{t}\bar{\mathrm{t}}}\xspace}

  \newcommand{\mT}{\ensuremath{m_{\mathrm{T}}}\xspace}
   \newcommand{\met}{\ensuremath{{E\!\!\!/}_{\mathrm{T}}}\xspace}
  
\begin{document}
\begin{titlepage}
\pubblock

\vfill
\Title{Single Top Quark Production Measurements in CMS}
\vfill
\Author{ Ferdos Rezaei Hosseinabadi\\
on behalf of the CMS Collaboration
}
\Address{\institute}
\vfill
\begin{Abstract}
The CMS experiment has measured the electroweak production of top
quark in different singly production modes: t-, tW- and s-channel. Recent
results on inclusive and differential cross section measurements of t-channel
at 13\TeV and of the fiducial cross section at 8\TeV are shown. The search
for single top quark production in s-channel at 7 and 8\TeV as well as the
search for associated production of Z boson and single top are presented. All
measurements are in agreement with the Standard Model prediction and
no sign of new physics is observed.
\end{Abstract}
\vfill
\begin{Presented}
$9^{th}$ International Workshop on Top Quark Physics\\
Olomouc, Czech Republic,  September 19--23, 2016
\end{Presented}
\vfill
\end{titlepage}
\def\thefootnote{\fnsymbol{footnote}}
\setcounter{footnote}{0}

\section{Introduction}

Top quark at the LHC is produced mainly in pairs through the strong interaction,
however it can be produced singly via the electroweak interaction. Based on the W
boson virtuality, single top production is categorized into t-, tW- and s-channel in the
descending order in terms of SM prediction for their cross section. %
In the SM, the associated
production of a Z boson and top quark is expected to happen at a low rate, however, it can be enhanced by BSM effects.
In this document, the latest results of single top quark production measurements in the CMS experiment\cite{CMS} at 7, 8 and 13 \TeV center-of-mass energy are reported.

\section{Single top t-channel cross section measurement}
Single top quark production in the t-channel has the highest predicted cross section amongst the foreseen production mechanisms. 
In the following, recent measurements on inclusive, differential and fiducial cross section of t-channel are reviewed.

\subsection{Inclusive cross section at 13\TeV}

The measurement of inclusive t-channel cross section at 13\TeV is performed using
2.3\invfb data collected in 2015 by  CMS\cite{Sirunyan:2016cdg}. This analysis is performed in the muonic decay channel when the W boson from the top quark decay, further decays to a muon
and a neutrino. Therefore, the final state is characterized by
the presence of exactly one isolated muon, missing energy, one b jet from the top quark
decay, and a light-flavour jet in the forward region. Events that contain exactly
one muon candidate with transverse momentum $\pt > 22\GeV$, pseudorapidity $|\eta| < 2.1$, and relative isolation $I_{rel} < 0.06$
are selected, where $I_{rel}$ is defined as the ratio of transverse energy sum deposited in a
cone with a size of $\sqrt{\delta \phi ^2 + \delta \eta^2} = 0.4$ around muon direction and its transverse momentum. Events with
additional muon or electron candidates 
are rejected. Reconstructed jets 
 are required to have $\pt > 40\GeV$ within $|\eta| < 4.7$. Selected jets can be further
categorized using a b-tagging discrimination algorithm in order to distinguish jets stemming from the hadronization of a b quark. 
Events are divided in categories named N-jet(s) M-tag(s) where N and M refer to the jet and b jet multiplicities. 
 The category
enriched in t-channel signal events is the 2-jets 1-tag, while the 3-jets 1-tag and 3-jets 2-tags categories are used to constrain the \ttbar contribution. 
In order to reduce the \QCD contamination, a cut on transverse mass of reconstructed W boson (\mT ) at
50\GeV is applied. 
The \QCD contribution  is extracted directly
from data by a fit to the whole range of \mT distribution and then extrapolation 
to $\mT > 50\GeV$. 
In this fit, the non-\QCD (\QCD) template is taken from MC simulation(data with inverse muon relative isolation $(I_{ rel} >0.12)$). 
After the dedicated event selection, a neural network is trained in the 2-jets 1-tag sample 
and  afterwards applied on all categories
which are further separated by the charge of the muon. To determine the
signal cross section, a simultaneous binned likelihood fit is performed
on the distributions of the multivariate discriminators. 
The measured cross sections for the top quark and top anti-quark production
are $\sigma_{\rm{t-ch.,t}} = 150 \pm 8(\rm{stat.}) \pm 9(\rm{exp.}) \pm 18(\rm{theo.}) \pm 4(\rm{lumi.})pb$ and $\sigma_{\rm{t-ch.,\bar{t}}} =
82 \pm 10(\rm{stat.}) \pm 4(\rm{exp.}) \pm 11(\rm{theo.})\pm 2(\rm{lumi.})pb$, respectively. The top quark to top anti-quark cross section ratio $R_{t-ch.}$ is measured in the fit
to be $R_{\rm{t-ch.}} = 1.81 \pm 0.18(\rm{stat.})\pm 0.15(\rm{syst.})$. 
The inclusive cross section is  used to determine
the absolute value of the CKM matrix element $|V_{tb}|$. Assuming $|V_{td}|$ and $|V_{ts}|$ 
to be much smaller than $|V_{tb}|$, 
 the  $|f_{LV} V_{tb}|$ is defined as the square root of the ratio of the measured cross section to
the SM prediction. It is measured to be $1.03\pm 0.07(\rm{exp.})\pm 0.02(\rm{theo.})$ 
where $f_{LV}$ is Wtb anomalous coupling form factor.

\subsection{Differential cross section at 13\TeV}

\begin{figure}[hpt]
 \begin{center}
         \subfloat[]{ 
           \includegraphics[width=0.4\textwidth]{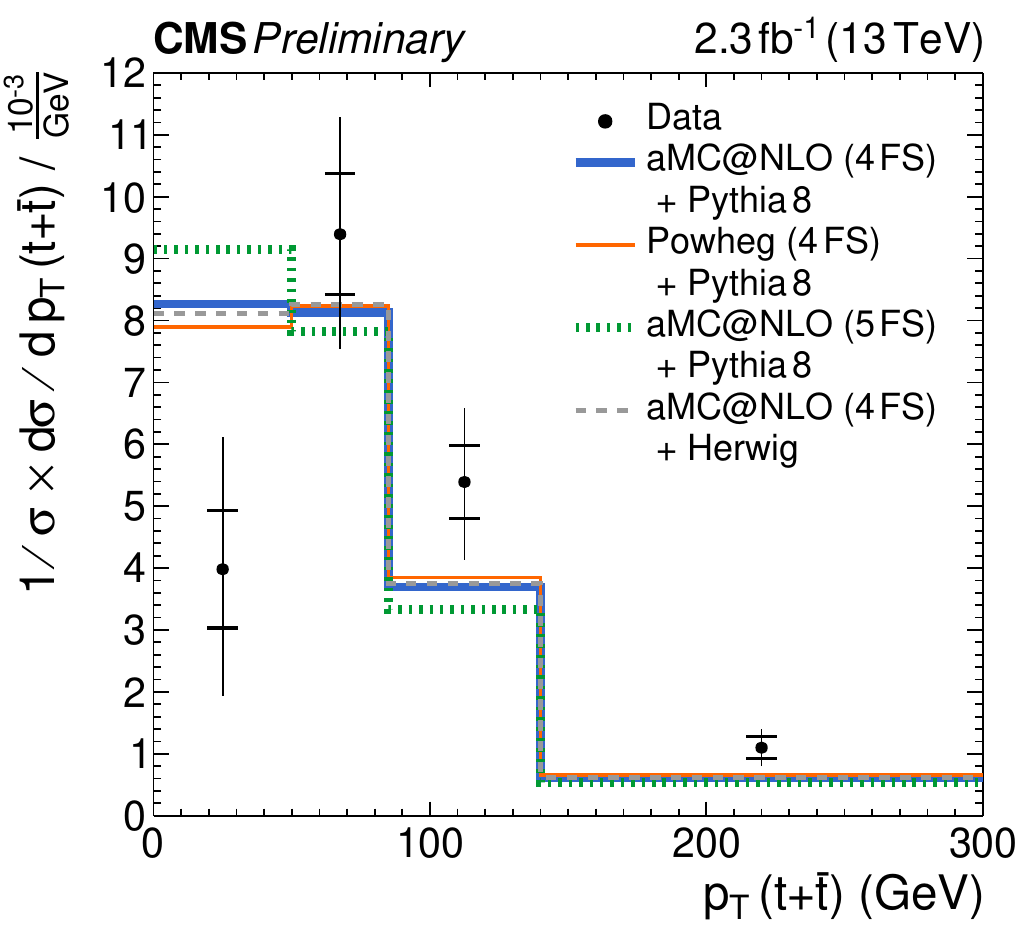}}
          \subfloat[]{   
             \includegraphics[width=0.4\textwidth]{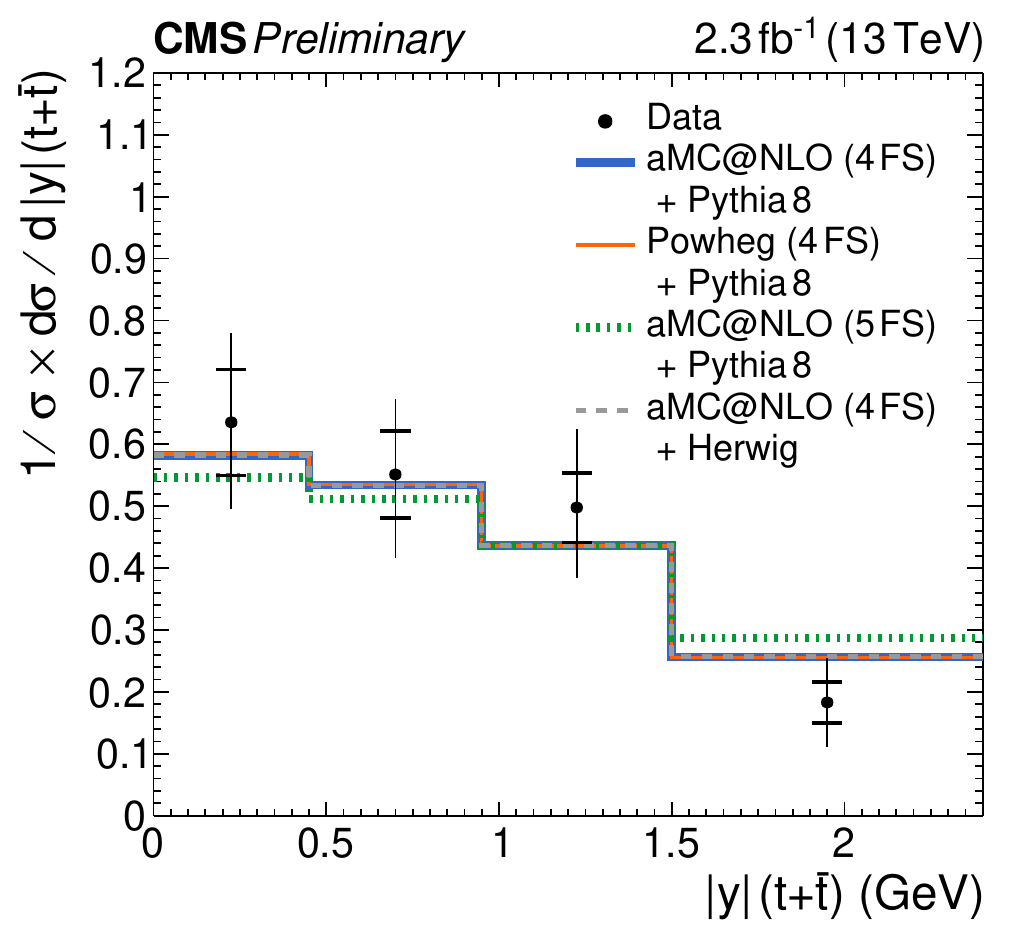}}
        \caption{  Measured differential cross section of t-channel as a
function of the top quark  transverse momentum (a) and  rapidity(b). The total uncertainty per bin is shown with  vertical bar.\label{fig2}}
        \end{center}
      \end{figure}

A differential cross section measurements of t-channel is performed using
2.3\invfb collision data at 13\TeV, collected during 2015 by the CMS experiment\cite{CMS:2016xnv}. Very close
event selection to the inclusive measurement is exploited to enhance  signal to background ratio. 
To further discriminate the main backgrounds 
from the signal events, a Boosted Decision Trees(BDT) is trained in 2-jets
1-tag and applied on all categories (2-jets 1-tag and the 3-jets 1-tag
and 3-jets 2-tags). 
Results are obtained using binned maximum likelihood fits separately
in each bin of the measurement. The likelihood is defined by taking the \mT distribution for the events with $\mT < 50\GeV$, and the BDT discriminant
otherwise. In this analysis, the estimation of \QCD backgrounds is performed
simultaneously in the final fit as the \mT used in likelihood definition  is a powerful
handle on the \QCD estimation. 
To achieve the differential cross section as a function
of the parton-level single top quark transverse momentum and rapidity (y), an
unfolding procedure is applied using the TUNFOLD algorithm \cite{Schmitt:2012kp}.\\ 
The measured  unfolded distributions of the top quark \pt and $|\rm y|$ normalized to the measured inclusive cross section of t-channel, are shown in Figure \ref{fig2}  for data and simulation.
The large uncertainty for the first \pt bin is due to low acceptance for signal
events and the large sensitivity to the systematic uncertainties.
\subsection{Fiducial cross section at 8\TeV}
A fiducial cross section measurement of single top in t mode is performed using 19.7\invfb data collected at 8\TeV\cite{CMS:2015jca}. 
In this analysis leptonic decays of top to an electron or muon are studied. Events with exactly one lepton are selected. Muons (electrons) are required to have  $\pt > 26 (30)\GeV$ within $|\eta|<2.1(2.4)$, and to be isolated  with 
$I_{rel}<0.12 (0.1)$. Jets with $\pt > 40 \GeV$ and $|\eta| < 4.7$ are selected. 
Events with two jets where one of them passes the  b-tagging criteria are used to extract the cross section. 
To reject \QCD events with no genuine  neutrino, a cut on \mT at 50(\met at 45) in muon(electron) channel is applied. The missing energy \met is defined as the magnitude of the vector sum of  all particles momenta in the final state.  Finally, the reconstructed top quark mass is required to be $130<\mt<220$. In a fiducial measurement, simulated events in 'generator' level are required to lie in the fiducial volume defined in the event selection in 'reconstruction' level.
In this analysis, the fiducial volume is defined with exactly one dressed lepton\footnote{lepton and its associated clustered radiations together} in $|\eta| < 2.4$ with $\pt>30\GeV$ and two jets with $\pt > 40\GeV$, one  identified as  b jet in presence of a b hadron in $|\eta| < 2.4$ and the other in $|\eta| < 5.0$.
The fiducial cross section is measured using number of observed events ($N_{obs}$):
\begin{eqnarray}
\sigma_{\rm{t-ch.}}^{\rm{fid}}& =& \frac{\rm{N_{obs}} }{ \epsilon ^{\rm{fid}}\mathcal{L}}, \nonumber 
\end{eqnarray}
where $\mathcal{L}$ is integrated luminosity, $\epsilon^{\rm{fid}}= N^{\rm{MC}}_{\rm{rec}}/N^{\rm{MC}}_{\rm{fid}}$, $N^{\rm{MC}}_{\rm{rec}}$ is number of events after full selection and 
$N^{\rm{MC}}_{\rm{fid}}$ is number of events in the fiducial phase space. The Fiducial cross section can be translated to total cross section by $\sigma_{\rm{t-ch.}} ={\sigma_{\rm{t-ch.}}^{\rm{fid}}}/{ \rm{Acc^{fid}}}$ where  $Acc^{\rm{fid}} = {\sigma_{\rm{t-ch.}}^{\rm{fid,MC}}}/{\sigma_{\rm{t-ch.}}^{\rm{MC}}}$ is the acceptance of the fiducial volume. 
The measured fiducial and total cross section in presence of experimental and theoretical 
uncertainties are:
\begin{eqnarray}
\sigma_{\rm{t-ch.}}^{\rm{fid,obs}} &=& 3.38 \pm 0.25(\rm{exp.}) \pm 0.06(\rm{scale})\pm 0.08 (\rm{PDF}) \pm 0.17(\rm{NLO-subtr.})pb,  \nonumber \\ 
\sigma_{\rm{t-ch.}}^{\rm{obs}} &=& 87.2 \pm 6.5(\rm{exp.}) ^{+0.2}_{-1.1}(\rm{scale})\pm 1.9  (\rm{PDF}) \pm 0.7 (\rm{NLO-subtr.})pb. \nonumber
\end{eqnarray}

\section{Single top s-channel cross section measurement}
Despite its very small cross section among single top at the LHC, s-channel  is of special interest.  Beside validation of SM, it is sensitive to  physics beyond the SM, e.g. models predicting $W^\prime$ boson or charged Higgs as mediator can enhance the cross section. 
A search for s-channel is performed in CMS using collision data corresponding to the integrated luminosities of 5.1 and 19.7\invfb at centre-of-mass energies of
7 and 8\TeV, respectively \cite{Khachatryan:2016ewo}. In this measurement, the leptonic decay of top quark  is  considered; at 7\TeV only  muon channel is studied  while at 8\TeV both the
muon and electron channels are included.
Reconstructed muons with $\pt > 20 (26)\GeV$ at 7(8)\TeV within the trigger acceptance ($|\eta| < 2.1$) are selected. Muons should be isolated with $I_{rel} < 0.12$ measured in a cone of size $\Delta R = 0.4$.   
Isolated electrons with $I_{rel} < 0.1$ in a cone of size $\Delta R = 0.3$, with $\pt > 30\GeV$ in $|\eta| < 2.5$ are selected. Jets within $|\eta| < 4.5$ and $\pt > 40\GeV $ are considered and a b-tagging algorithm identify the b jets .
 Three statistically independent  event categories are defined:   2-jets 2-tags as the s-channel enriched sample, 2-jets 1-tag category for controlling the t-channel and \wjets backgrounds and  3-jets 2-tags category to constrain \ttbar as the dominant background.
In  2-jets 2-tags another selection is applied on  number of jets with $20 < \pt< 40 \GeV$ to further reject \ttbar background. 
In the 7\TeV analysis,  the \mT distribution is used to estimate \QCD multijet contribution. 
But in the 8\TeV analysis, BDT discriminants 
 are used to  estimate and reject multijet events. 
 The BDT distribution is used to define a \QCD enriched region. Maximum likelihood fits are performed in \QCD enriched regions and results are extrapolated outside. The \QCD template is taken from data with non-isolated lepton. \\
For signal extraction, other  BDTs are trained per each category in electron and muon channels at 7 and 8\TeV and maximum likelihood fit is employed to measure the cross section. 
The cross section is measured to be $7.1 \pm 8.1 (stat + syst) pb$ at 7\TeV and $13.4 \pm 7.3 (stat
+ syst) pb$ at 8\TeV, corresponding to a combined signal rate relative to SM expectations of
$2.0 \pm 0.9 (stat + syst)$.  The combined signal strength is measured with the observed significance of 2.5 standard deviations with 1.1 standard deviations expected. The observed and expected upper  limits at 95\% CL on the combined signal strength are found to be 4.7 and 3.1. 
\section{Associated production of Z boson and single Top at 8\TeV}
Having high centre-of-mass energy and luminosity at the LHC enhances the chance for observing rare SM or BSM single top quark production processes. A search for the SM production of a single top quark in association with a Z boson is performed at 8\TeV\cite{CMS:2016bss}. 
 In this process, the top quark is produced via the t-channel mechanism and the Z boson
is radiated off one of the initial or final state quarks or from the exchange of W bosons.
Therefore,  considering leptonic decay of top quark  and Z boson, the final  state consists of a  neutrino, one  b quark and three leptons and a quark. 
Each selected event contain exactly three leptons each with $\pt>20\GeV$ and $|\eta|<2.5 (2.4)$ for electrons(muons). The electrons(muons) should be isolated, $I_{rel} <0.15(0.12)$ measured in cone of size 0.3(0.4).
There are four possible leptonic combinations in the final state: $eee, \mu \mu \mu, \mu \mu e $ and $ee \mu$. 
To fulfil the criteria of Z boson decay,  two same flavour leptons are required to have opposite charges and invariant mass around Z mass ($76<m_{ll}<106 \GeV$).  
In $eee, \mu\mu\mu$ channels, the pair with closest mass to Z peak mass is selected for Z boson reconstruction. The third lepton is used to reconstruct the W boson. 
Jets with $\pt>30\GeV$ within $|\eta|<4.5$ are selected. Event is accepted if it contains two or more jets while one or more pass b-tagging criteria. To further reject backgrounds, a cut on $\mT$ is applied at 10\GeV. In addition to signal region defined by the above criteria 
a  control region enriched with DY and WZ processes is defined with the same selection but vetoing events containing b-tagged jet. 
A BDT is trained in signal sample to separate signal events from background processes. 
The signal extraction is performed using a simultaneous likelihood fit on distributions of BDT output in the signal region and \mT in the background-enriched control region in four different leptonic final states.  
 Template for backgrounds with non-prompt leptons (\ttbar and  DY) is taken from non-­isolated sample.\\ 
The combined measured signal cross section is found to be $\sigma_{tZq \rightarrow \ell\nu b \ell\ell q } = 10 ^{+8}_{-7} fb $ which is in agreement with the SM prediction of $8.2 ^{+0.59} _{-0.03} (\rm{scale})$ $fb$. The signal is measured with the observed and expected significances of 2.4 and 1.8, respectively. An upper limit on the  $tZq$ cross section is set at 21 $fb$ at the 95\% CL.

\end{document}